\documentclass[%
 ,twocolumn%
,amssymb, amsmath,nobibnotes, aps, prl]{revtex4}

\usepackage{graphicx}
\usepackage{dcolumn}

\begin{document}

\begin{abstract}
We formulate and apply a continuum model that
incorporates elasticity, yield stress, plasticity and viscous drag. It is
motivated by the two-dimensional foam rheology
experiments of Debregeas \emph{et al.}
[G. Debr\'egeas, H. Tabuteau, and J.-M. di Meglio, {\em Phys. Rev. Lett.}
{\bf 87}, 178305
(2001)] and Wang \emph{et al} [Y. Wang, K. Krishan, and M. Dennin, \emph{Phys. Rev. E} {\bf 73}, 031401 (2006)],
and is successful in exhibiting their principal features an exponentially decaying velocity profile and strain localisation. Transient effects are also identified.

\end{abstract}

\title{Two dimensional foam rheology with viscous drag}
\author{E. Janiaud}
\email{janiaude@tcd.ie}
\author{D. Weaire}
\author{S. Hutzler}
\affiliation{School of Physics, Trinity College Dublin, Ireland}
\maketitle


While initially two dimensional (2d)
foams were introduced only as a simple model system for numerical and
theoretical studies \cite{Bolton-rigidityloss, Durian-bubblePRE},
recent years have also seen a variety of rheological experiments on
so-called quasi 2d foams, \emph{i.e.} foams consisting of a single layer
of bubbles \cite{Vaz-quasi2D, debregeas1, Janiaud-JFM, Cantat, Lauridsen-prl2004, Dennin-comparedraft}. 
Using bubbles trapped between two glass plates (Hele-Shaw cell)
in a cylindrical Couette geometry (the foam is contained between two concentric cylinders), Debr\'egeas \emph{et al.} found that the flow of the foam \emph{localises} near the inner moving wall with an exponential velocity profile forming \emph{shear-bands} \cite{debregeas1}. While quasi static cellular simulations \cite{debregeas2,cox05}, showed some agreement with the results, they continue to excite debate \cite{Lauridsen-prl2004}, especially as regard
to the localisation of shear and deformation \cite{Janiaud-JFM}, which is
the salient feature of the experiment. Recently, Wang \emph{et al.} have
extended shear experiments to the simpler planar geometry
\cite{Dennin-comparedraft}. While their experiments using bubbles between a
liquid pool and a glass plate showed the formation of shear bands with an
exponential velocity profile, a nearly linear velocity profile was
obtained for bubble floating on the liquid (bubble raft or Bragg raft). 
This has evidenced the primordial role played by the method used to confine the bubbles and indicates that the non-uniform stress imposed by the Couette geometry is not sufficient to explain the formation of shear bands with exponential decaying velocity.

In this paper, we introduce an elementary continuum model for the analysis
of rheological properties of a two-dimensional foam. It includes a viscous drag that has no counterpart in conventional 3d foam rheology. Our model is therefore closely related to the \emph{2d viscous froth model} \cite{Kern2004-ViscousFroth} which was designed to enable dynamic simulations to be undertaken with the full
cellular structure of the foam, and included just such a viscous drag.
Here the viscous drag will enter as a term in the continuum description,
depending on a local \emph{average} of the boundary velocity. Experiment
and theory have already addressed this force as it arises in the flow of
bubbles in cylindrical tubes and in narrow channels \cite{Cantat}.  It is
often associated with the name of Bretherton who showed that the force
varies with the two-thirds power of velocity \cite{Bretherton}.  In some
circumstances, a power law of one-half is suggested \cite{Denkov}.  
Nevertheless, as in the case of the 2d viscous froth, we adopt a linear
form in order to keep the model and the analysis simple, in a search for
qualitative and semi-quantitative understanding. In other respects, the
model is akin to the familiar Bingham model of a substance that has a
yield stress \cite{book} and an internal viscosity. This, or one of its
variants, is often invoked in the analysis of bulk foams.  However, as in
the recent work of Takeshi and Sekimoto \cite{Sekimoto2005}, we also
include an elastic response, so that the model we propose has four key
ingredients: elasticity up to a yield stress, plasticity, internal
viscosity and a viscous drag force.

\begin{figure}
	\center
\includegraphics[width=0.7\linewidth]{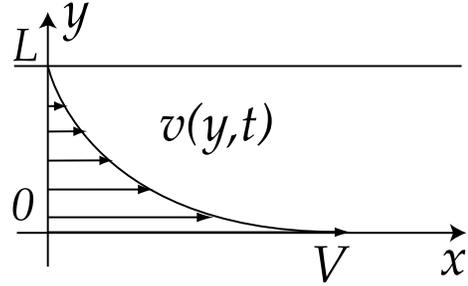}
	\caption{Geometry of the case considered here in which the 
velocity $v(y,t)$ and the displacement $u(y,t)$ are functions of the 
$y$ coordinate and time $t$.
}
	\label{fig1}
\end{figure}

 While it is amenable to obvious generalisation, the model will be defined here for the simple planar shear geometry, as in \cite{Dennin-comparedraft}.
Displacement $u(y,t)$ and the velocity $v(y,t)=\frac{\partial u(y,t)}{\partial t}$ are in the $x$ direction only, as when shear takes place between two parallel infinite
boundaries in that direction (see figure \ref{fig1}).  This reduces the problem to one dimension. 
Strain and strain rate are
reduced to scalars $\gamma(y,t) = \frac{\partial
u}{\partial y}$ and $\dot \gamma(y,t)= \frac{\partial \gamma}{\partial t}$.

We will neglect inertia throughout, so that the total force acting on an
element of fluid at $y$ must be zero. 
Forces arise from the gradient of the shear stress $\sigma(y,t)$
and the drag force per unit area, $F=-\beta v$, where $\beta$ is the mean drag
coefficient. In two dimensions, stress has the dimension of a force
divided by a length and $\beta$ is expressed in units of
force $\times$ time per volume.  The required force
balance is: 
\begin{equation} 
\frac{\partial \sigma}{\partial y} = \beta v.
\label{friction} 
\end{equation}

It remains to specify the constitutive relation for $\sigma$ in terms of $\gamma$ and $\dot \gamma$. For simplicity, we capture the desired ingredients of elasticity, yield stress and plasticity with the following relation: 
\begin{equation}
\sigma = \sigma_{Y} f \left( \gamma/\gamma_{Y} \right) + \eta \dot \gamma.
\label{constitutive} 
\end{equation} 
Here, $\sigma_{Y}$ is the yield stress and $\gamma_{Y}$ is the yield strain.  We choose 
$f(\gamma/\gamma_{Y})= \tanh(\gamma/\gamma_{Y})$
which roughly corresponds to a typical  2d static stress-strain relation for
foams \cite{book}. 
For foams $\gamma_Y$ is of the order of unity and we shall set it equal to unity here. The final term in eqn. \ref{constitutive} is the usual
strain-rate term of the Bingham model. Note that for foams, the viscosity $\eta$ depends on the strain $\gamma$. 
For low strain, the dissipation is due to the stretching of films and occurs at the same rate than the applied deformation. For high strain, it is mainly due to topological changes which leads to the disappearance and creation of films. This occurs at much higher rate than the applied deformation \cite{Prisme-Hutzler}. Nevertheless, assuming a constant viscosity is helpful in our elementary model.
 A very important
restriction requires that eqn. \ref{constitutive} is used only when the
strain rate $\dot \gamma$ always has the same sign (negative
in what follows), which is the case in the experiments we are referring to.
In further work we will include hysteretic effects, which are
very important, but for now we accept this restriction.

We can non-dimensionalise equations \ref{friction} and \ref{constitutive} 
by introducing the natural length scale $L_{0} = (\eta/\beta)^{1/2}$ and 
natural time scale $T_{0} = \eta/\sigma_{Y}$. From now, length and time will be expressed in units of $L_{0}$ and $T_{0}$.
A convenient representation of eqns.
(\ref{friction})
and (\ref{constitutive}) is
\begin{equation}
\frac{\partial^{2} v}{\partial y^2} - v   =  - \frac{\partial}{\partial
y}f\left(\frac{\partial u}{\partial y}\right) ,
\label{velocity1}
\end{equation}
where 
\begin{equation}
v  =  \frac{\partial u}{\partial t}.
\label{velocity2}
\end{equation}

The model can be solved analytically in various cases and limits. More
generally, a numerical scheme of integration can be used to follow the
time dependence of the variables, as follows. We discretise $y$ and $t$ with small steps $\Delta y$ and $\Delta t$, using lowest order expressions for derivatives. Given a knowledge of $u$ in steps
up to time $t$, $\frac{\partial u}{\partial t}$ may be estimated as a
backward derivative and eqn. (\ref{velocity1}) may be solved for
$v(y,t)$ with the imposed boundary conditions. Eqn. (\ref{velocity2})
then enables us to update  $u$ to $t + \Delta t$. (In practice an  
Improved Euler method was used for the integration in time.)

We will only consider the case in which the boundary at $y=0$ is given 
a finite velocity $V$ for all time $t$, while the boundary at $y=L$ is
held fixed. Correspondingly, $u(y=0,t)=Vt$ and $u(y=L,t)=0$.
For the results presented here, we set $L=15$ and $V$ takes various values. 
The quantity $\Gamma= V t/L$ may be regarded as the total applied shear at
time $t$. 

\begin{figure}
	\center 
\includegraphics[width=1.0\linewidth]{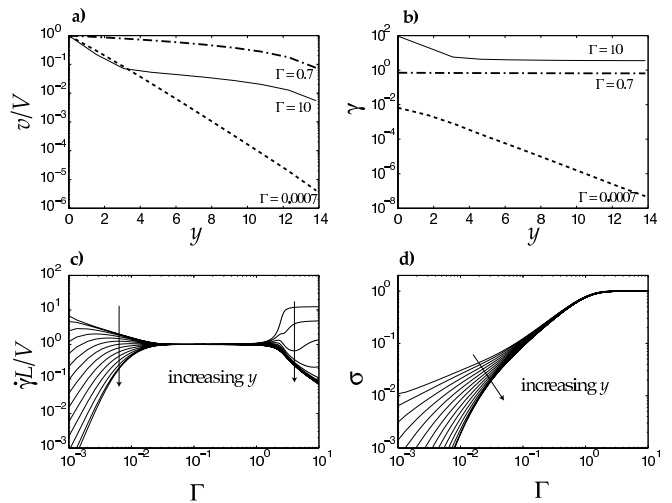}
	\caption{
(a,b) Profiles of velocity $v/V$ and magnitude of the strain $\gamma$ for three different times, represented by the total applied shear $\Gamma=V t/L$, shown in semi-log scale. This exemplifies three regimes of exponential/linear profiles. The regimes also feature in (c), the variation of the magnitude of strain rate $\dot \gamma$ and (d), the variation of the magnitude of stress $\sigma$ with total applied shear $\Gamma$, shown in log-log scale.
In all the calculations shown we have chosen $L=15$ and a low boundary
velocity of $V=0.005$ at $y=0$.
} 
\label{fig2} 
\end{figure}

The numerical results presented in Figure 2 are for low velocities
$V\ll 1$ and show the existence of several regimes as the total applied
shear is increased.
                                                                                
Regime I is observed for small total applied shear, at which both velocity
and strain profiles (Figures \ref{fig2} a and b) are close to exponentials.  Regime II
is characterised by a linear velocity profile and a homogeneous strain. In regime
III both velocity and strain profiles combine an exponential decay close
to the moving boundary and a linear decay close to the fixed boundary.
With further increase of total applied shear the linear tails diminish,
leading to an asymptotic steady state (regime IV) similar to that for
small applied shear.

The existence of these distinct regimes is also evident from the plots of
strain rates and stress as a function of total applied shear as shown in
Figures \ref{fig2} (c) and (d) respectively. While regime I is characterised by a
strong localisation of both strain-rate and stress, in regime II
($10^{-2}<\Gamma<1$) the strain-rate is homogeneous. The asymptotic
steady state of regime IV is again characterised by strong localisation of
the strain rate. Stress is saturating to its maximal magnitude $1$ for all
values of $y$.

\begin{figure}
	\center
\includegraphics[width=0.42\linewidth]{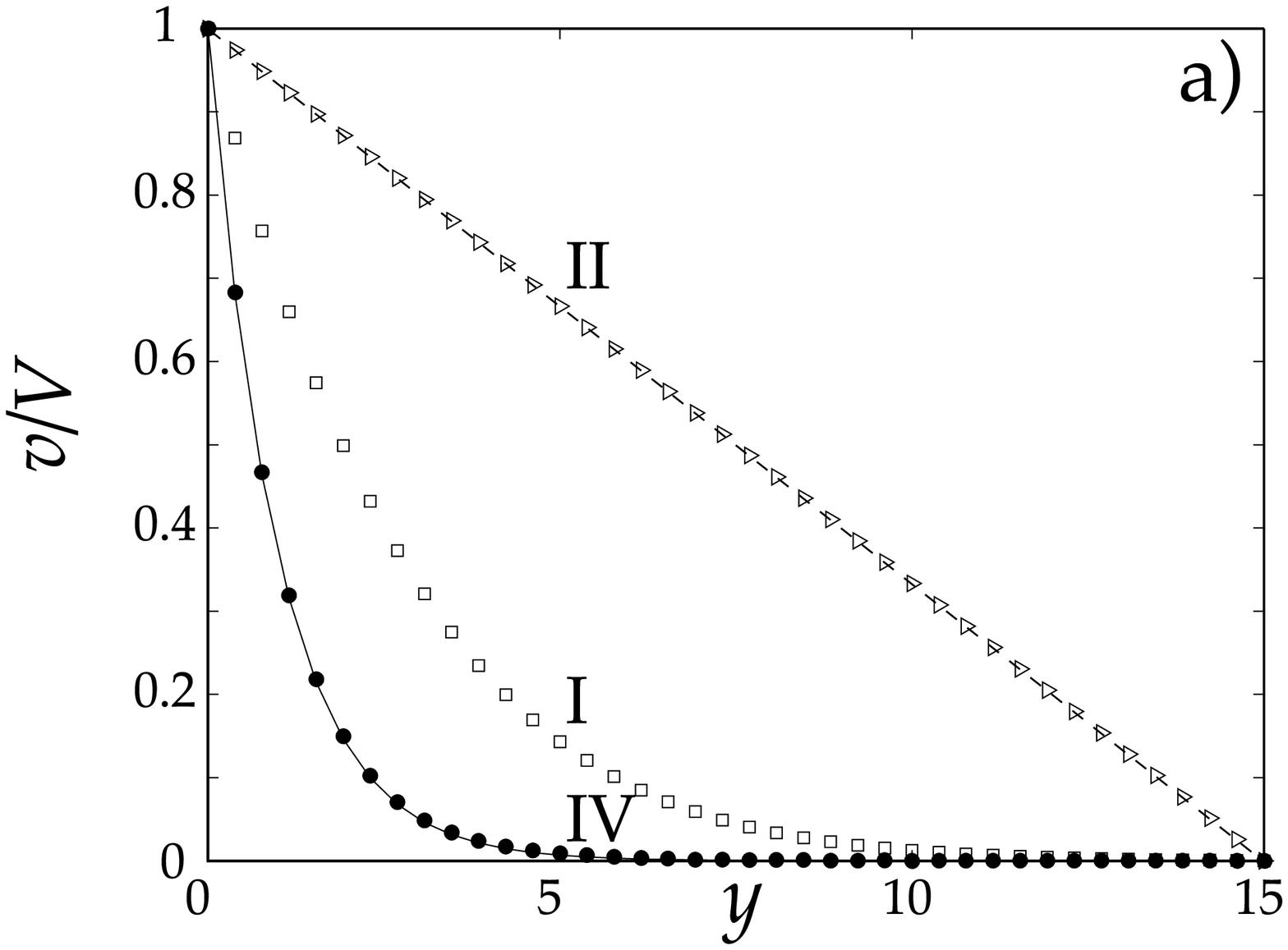}
\includegraphics[width=0.42\linewidth]{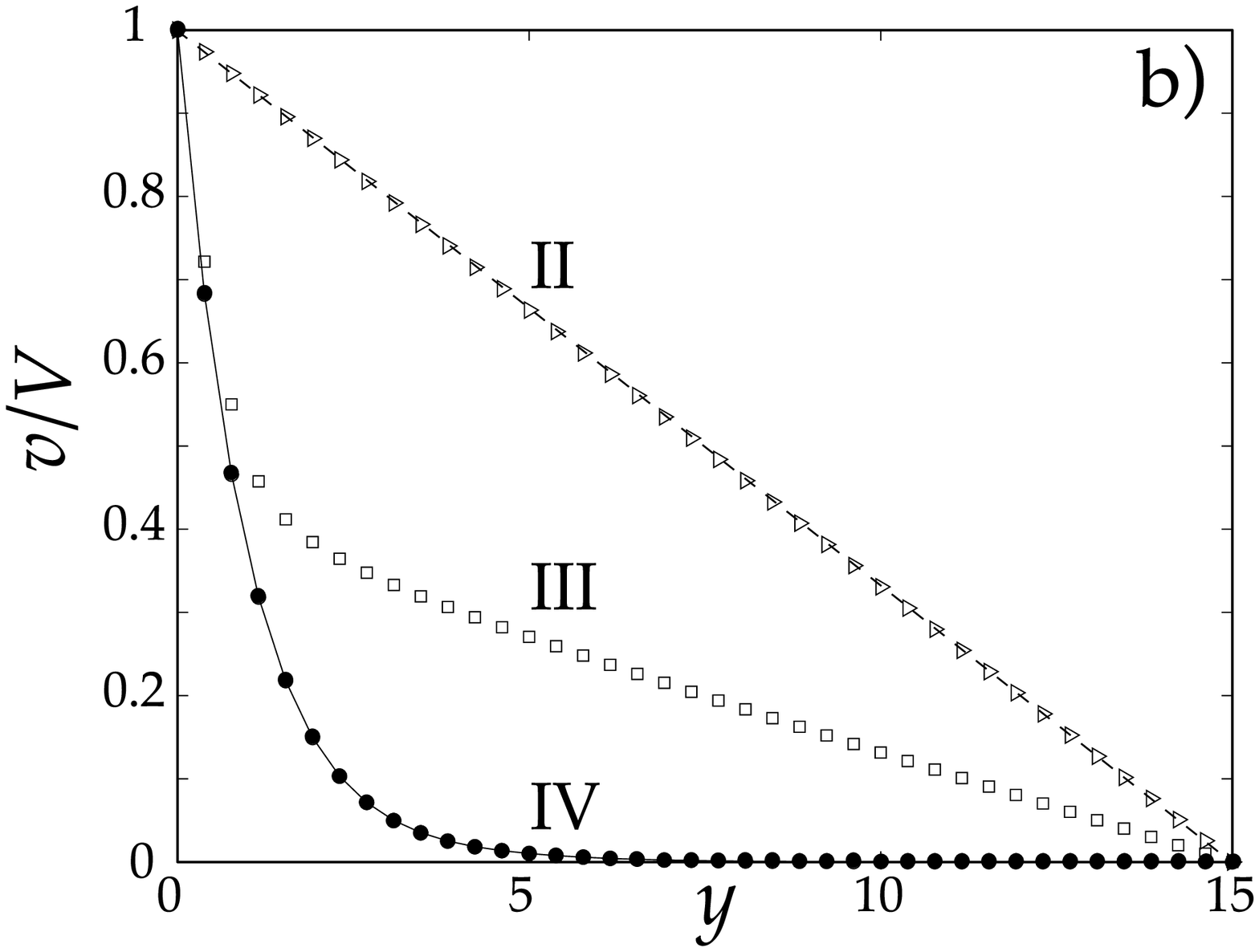}
\caption{Velocity profiles scaled by $V$ for $L=15$ (numerical results for $\square$ $V=0.001$, $\triangleright$ $V=0.3$ and $\bullet$ $V=60$). 
(a) is for total applied shear $\Gamma = 0.1$ and shows the succession of regime II, regime I and finally regime IV. 
(b) corresponds to the transition from regime II to III to IV as obtained for $\Gamma=1$.
The dashed line represents the linear solution given by eqn. (\ref{vlin})
and the solid line is the steady states solution given by eqn. 
(\ref{vsteady}) corresponding to the exponential localisation.
}
\label{fig3}
\end{figure}

\begin{figure}
	\center
\includegraphics[width=1\linewidth]{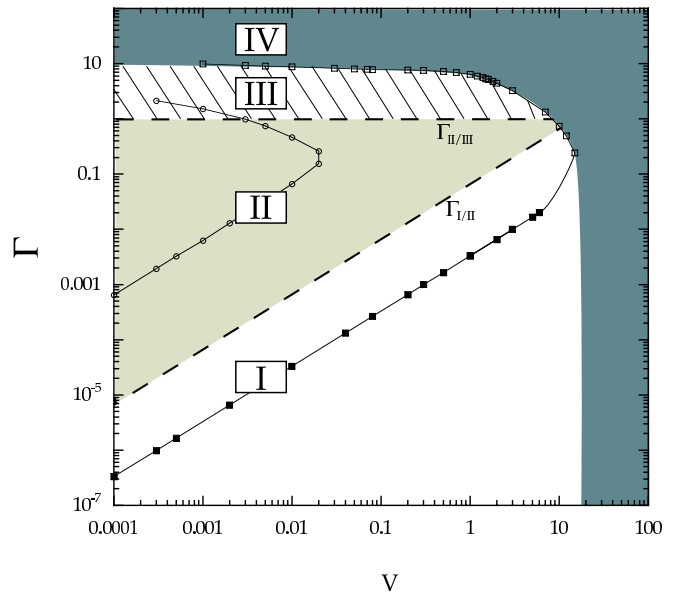}
	\caption{
Qualitatively different velocity profiles are found in different regions of the
$\Gamma-V$ diagram. Regime I: exponential. Regime II linear. Regime
III: combined exponential/linear. Region IV: approach to final steady state
velocity profile with an exponential localisation. The boundaries $\Gamma_{I/II}$ and $\Gamma_{II/III}$ (dashed lines) are defined as in the text.
The area to the left of the computed $\circ$  data points is defined so that the relative error between the velocity profile and the linear profile characterizing Regime II (eqn. \ref{vlin})  is smaller than 1\%.
A similar 1\% threshold based on equation \ref{vsteady} has been used as a numerical criterion for the computation of the boundary between regime III and IV ($\square$). This threshold is also used for the computation of the $\blacksquare$  data points.
}
\label{fig4}
\end{figure}

Figure \ref{fig3} shows the velocity profiles obtained for the same total
applied shear but for different shearing velocity $V$. For $V \ll 1$, the velocity varies linearly, corresponding to the regime II.
For $V \gg 1$, the profile approaches the asymptotic form (regime IV). 
For  $V \approx 1 $, we can have either regime I (for $\Gamma = 0.1$ on figure \ref{fig3} a) or regime III (for $\Gamma = 1$ on figure \ref{fig3} b) where we see an initial 
exponential decay followed by a linear tail (regime III).
Figure \ref{fig4} represents the different regimes encountered on a semi-quantitative $\Gamma-V$ diagram. Depending on the shearing velocity $V$, several scenarios are possible before reaching the steady state of the regime IV. 

In order to understand these features, we return to the governing eqn.
(\ref{velocity1}), and reduce it by various approximations.
For small time (regime I), $u$ is small and we neglect the right hand side of
eqn.(\ref{velocity1}) which is approximately $-\frac{\partial^2 u}{\partial y^2}$. The remaining equation
\begin{equation}
\frac{\partial^{2} v}{\partial y^2} - v = 0,
\label{e:steady}
\end{equation}
has the elementary solution
\begin{equation}
v=V\frac{\sinh(L-y)}{\sinh(L)}.
\label{vsteady}
\end{equation}
Note that this solution does not vary with time, implying that the system
jumps instantaneously to the above velocity profile. This is indeed
consistent with what is found in the numerical treatment, and is a
consequence of the singular initial condition and the neglect of inertia.
Provided $L>>1$ in the reduced units, this solution is approximately an exponential
over most of the range. The exponential profile survives until
$\frac{\partial^2 u}{\partial y^2}$ becomes large, and overtakes the term
proportional to $v$.

Neglecting the term proportional to $v$ in eqn. \ref{velocity1}, rather than that on the right hand
side, and approximating the latter as already stated, we obtain
\begin{equation}
\frac{\partial^{2} v}{\partial y^2} = - \frac{\partial ^{2} u}{\partial y^2}. 
\label{e:linear_regime}
\end{equation}
Hence $u+v = a(t)y+b(t)$. Writing $v=\frac{\partial u}{\partial t}$, and integrating again gives $u = A(y)e^{-t} + a(t)y +b(t)$.
This shows that the solution which develops after some time is linear, with
a decaying transient part. The decay time is unity, in the units used.
Applying the boundary conditions at $y=0$ and $y=L$, the linear variation of the velocity is then given by 
\begin{equation}
v = V\left(1- \frac{y}{L} \right),
\label{vlin}
\end{equation}
in excellent agreement with the simulation (see fig. \ref{fig3}).

A further transition (regime III) takes place when the approximation $\tanh z \sim z$ fails, and can be replaced by $\tanh \sim 1$, 
as the strain $\gamma$ increases beyond
the yield strain $\gamma_Y$. 
At any given time in this regime, the second
approximation replaces the first for $y>y_0$. Thus the same exponential of regime I is to be expected for $y<y_0$, continued by the linear solution of regime II for $y>y_0$. 

As the time $t$ tends to infinity (regime IV), $y_{0}$ tends to $L$
and the solution returns to the effectively exponential form of eqn.(\ref{vsteady}). The simulations are in excellent agreement with this profile, as shown by the solid line in figure \ref{fig3}. 
This solution is only asymptotically reached  because the strain in the vicinity of the fixed wall tends to zero due to the strong localization that appears close to the moving wall. A closer analysis of this approach is possible but will not be pursued here: suffice it to say that it is a slow (power law) convergence.

The boundaries between these regions may be identified as follows.
That between regimes I and II may be found by comparing the magnitudes of
the terms neglected in their respective approximations. Using the solutions given in eqns. (\ref{vlin}) and (\ref{vsteady}), gives $\Gamma_{I/II}\sim V/L$,
in agreement with the linearity in $V$ found in numerical computation of the I/II boundary (see fig. \ref{fig4} ).  
 Similarly, we enter regime III when the maximum value of strain $\gamma$
reaches $\gamma_Y$, which for the linear solution occurs at
$\Gamma_{II/III}\sim 1$, which is in reasonable agreement with the numerical data shown in fig.\ref{fig4}. Putting these together, we see
that regime II is eliminated entirely for $V > L$ in dimensionless units.
Reinstalling physical units this corresponds to a shear rate which exceeds
$\sigma_{y}/\eta$.

 The steady state obtained at high applied shear $\Gamma$ offers a very elementary candidate for the explanation of the phenomenon of localisation with exponential velocity profiles in 2d foams \cite{Dennin-comparedraft, debregeas1}. Our results allow for a direct comparison with the planar shear experiment on 2d foams of ref. \cite{Dennin-comparedraft}.
As eqn. (\ref{vsteady}) can be approximated (in
physical units) by $v/V \approx \exp(- y/L_{0})$, the velocity
measurements provide us with a direct determination of $L_{0}=(\eta/\beta)^{1/2}$. Expressed in
units of bubble diameter $d$, they correspond to $L_{0} \approx d $ for
the bubbles trapped between a glass plate and a pool of liquid. Our model also explains why no exponential localisation is found in the experiments using bubble rafts. In this case, the mean drag coefficient $\beta$ is expected to be very small, since there are no rigid plates, but rather the foam slides on underlying liquid. The decay length of the exponential
which scales like $(\eta/\beta)^{1/2}$ increases up to a value of the same order of magnitude than the system size $L$ and the velocity profile appears to be very close to a linear form. From eqn. \ref{vsteady} we also see, consistent with the experiments
of \cite{Dennin-comparedraft}, that in the steady state regime the scaled
velocity profiles $v/V$ do not depend on the shear rate $V/L$. 

Clearly the model can be applied more generally, for example to the circular Couette geometry. 
This suggest the use of polar coordinates $(r, \theta)$ which leads to an extra term $\sigma/r$  in the divergence of the stress of eqn.(\ref{friction}). Although a full mathematical treatment is required to solve the problem in the general case, it is possible to use our present results, provided  this extra term is much smaller than the viscous drag term $\beta v$. 
Assuming that the stress is dominated by the viscous contribution $\eta \dot \gamma$ during the steady state, one find that eqn. (\ref{friction}) still holds if the distance between the two cylinders is much bigger than $L_0$. This is the case when bubbles are confined in a Hele-Shaw cell where the velocity profiles are found to be exponential with $d < L_{0} < 2d$ \cite{debregeas1}. On the contrary, for a bubble raft sheared between two concentric cylinders, the velocity profile is not exponential but rather discontinuous \cite{Lauridsen-prl2004}. This can be explained by a viscous drag to small too overcome the non uniform stress effect of the Couette geometry.

We thus have seen that the exponentially decaying velocity profile in 2d foams, which is the signature of shear bands, is due the viscous drag generated by the bubbles on the confining plate. In due course more realistic forces (e.g. the Bretherton form) may be required, at
the expense of the extreme simplicity of what we have shown here. Most of the qualitative conclusions are likely to remain intact.

Our work was supported by the European Space Agency (MAP
AO-99-108:C14914/02/NL/SH, MAP AO-99-075:C14308/00/NL/SH)
and Science Foundation Ireland (RFP
05/RFP/PHY0016).


\end{document}